\documentclass{neuthist18}

\def\be{\begin{equation}}
\def\ee{\end{equation}}
\def\bea{\begin{eqnarray}}
\def\eea{\end{eqnarray}}

\newcommand{\Photo}{}

\begin{document}
\vspace*{4cm}
\title{THE MIKHEYEV-SMIRNOV-WOLFENSTEIN (MSW) 
EFFECT~\footnote{Talk given at {\it the International conference
on History of the Neutrino}, 
September 5 - 7, 2018, Paris, France}}

\author{ A.Y. SMIRNOV }

\address{Max-Planck-Institute, Theoretical Physics, Saupfercheckweg 1,\\
69117 Heidelberg, Germany}

\maketitle\abstracts{
Developments of main notions and concepts 
behind the MSW effect (1978 - 1985) are described.  They include  
(i) neutrino refraction, matter potential,  
and evolution equation in matter, (ii) mixing in matter,  
resonance and level crossing; (iii) adiabaticity condition 
and adiabatic propagation in matter with varying density. 
They are in the basis of 
the resonance enhancement of oscillations in matter with constant 
(nearly constant) density,  and  
the adiabatic flavor conversion in matter with 
slowly changing density. The former is realized in matter 
of the Earth and can be used to establish neutrino mass hierarchy.
The latter provides the solution to the solar neutrino problem 
and plays the key role in transformations of the supernova neutrinos. 
%%Further developments after initial phase from 1978 to 1986 
%%are outlined.
}

\section{Introduction}
%%%%%%%%%%%%%%%%%%%%%%%%%%%%%%%%%%%%%%%%%%%%%%%%%%%%%%

The MSW effect is  the (adiabatic) flavor conversion 
of neutrinos  driven by the change of mixing 
in the course  of propagation 
in matter with varying density. 
Here is my story,  
%%(a contribution to the history of MSW) 
%%from my (subjective) point of view, 
the way I learned,  understood and elaborated things. 
%%[[to conclusion]] This is contribution to the history of the MSW effect, 
%%the history would require more explorations and balanced result of different views.  
%%Historian  sthould collect stories from different 
%%partcipants to give more balanced description. 
I will describe the  main notions and concepts behind MSW. Essentially, they were     
developed in the period 1978 -  January 1986 (the time of Moriond workshop in Tignes). 
%%Not many papers have been published before 1986, 
%%just few talks were given and influence of different 
%%studies can be easily traced. 
%%It was explosion in number of papers after Moriond 86. 
I would divide this period in three parts: 

\begin{itemize}

\item
1978 - 1984:  Wolfenstein's papers and follow-ups.

\item
1984  - 1985: the Mikheyev-Smirnov mechanism.  

\item
1985 - beginning of 1986: Further developments.  

\end{itemize}
These items compose the outline of my talk.

%%%%%%%%%%%%%%%%%%%%%%%%%%%%%%%%%%%%%%%%%%%%%%%%%%%%%%%%%%%%%%%%%%%%%%%%%%%
\section{Wolfenstein's papers and follow-up}
%%%%%%%%%%%%%%%%%%%%%%%%%%%%%%%%%%%%%%%%%%%%%%%%%%%%%%%%%%%

I met Wolfenstein several times. 
Probably the last one was in St. Louis in 2008 at the 
dinner organized on occasion of my Sakurai prize. 
We were talking about the LBL and underground 
physics program in US. We never discussed MSW and its history.

\subsection{Neutrino oscillations in matter}
%%%%%%%%%%%%%%%%%%%%%%%%%%%%%%%%%%%%%%%%%%%%%%%%%%%%%%%%%%%%%%%%%%%%%%

Lincoln Wolfenstein (1923 - 2015)  
was 55 years old in 1978 when  the major results of his scientific 
life were obtained and the paper 
``Neutrino oscillations in matter'' \cite{w78a} was  published.   
This is a very rare case for a theorist. 
Less rare was that Wolfenstein's motivations and the main-stream results 
turned out to be not quite correct or relevant, while a lateral  branch, 
not appreciated by the author,  led to major developments. 

%%Wolfenstein's motivation was oscillations of massless neutrinos 
%%in models with hypothetical non-diagonal neutral currents (FCNC).  

That was the epoch of the neutral currents (NC) discovery 
and Wolfenstein's primary interest was in  
oscillations of massless neutrinos 
in model with hypothetical non-diagonal neutral currents (FCNC). 
In the paper  \cite{w75} written in 1975  Wolfenstein considered 
the NC neutrino interactions described by the Hamiltonian \cite{w75}
\begin{equation}
H=\frac{G_F}{\sqrt{2}} L_\lambda J^\lambda + h.c., 
\label{eq:hamilton}
\end{equation}
where the neutrino current is
\begin{equation}
L_\lambda  = \cos^2\alpha [\bar{\nu}_a 
\gamma_\lambda(1 + \gamma_5){\nu}_a 
+ \bar{\nu}_b \gamma_\lambda(1 + \gamma_5) 
{\nu}_b] +
\sin^2\alpha [\bar{\nu}_a 
\gamma_\lambda(1 + \gamma_5){\nu}_b 
+ \bar{\nu}_b \gamma_\lambda(1 + \gamma_5) {\nu}_a].      
\label{eq:currents}
\end{equation}
Here ${\nu}_a$ and ${\nu}_b$ are the neutrino states defined by the
charged current interactions and $\alpha$ is free parameter which fixes the relative 
strength of FCNC.  The scatterer's current is 
\begin{equation}
J_\lambda = g_p \bar{p} \gamma_\lambda p + 
g_n \bar{n} \gamma_\lambda n + 
g_e \bar{e} \gamma_\lambda e. 
\label{eq:currentscat}
\end{equation}

Due to the assumed symmetry $\nu_a \leftrightarrow \nu_b$  the NC interactions 
are diagonalized by the states 
\begin{equation}
\nu_1 = \frac{1}{\sqrt{2}} (\nu_a + \nu_b), ~~~~
\nu_2 = \frac{1}{\sqrt{2}} (\nu_a - \nu_b).  
\label{eq:eigen}
\end{equation}
The extreme case is purely off-diagonal NC, $\cos^2\alpha = 0$, 
in which ``neutrinos were never the same''. 
In interactions they change their flavor completely. 
A possibility to test this model was the central objective 
of the paper~\cite{w78a}, and the main notions were elaborated in this  framework.  \\

%%1. Neutrino oscillations in matter

{\it 1. Refraction of neutrinos.}
%%%%%%%%%%%%%%%%%%%%%%%%%%%%%%%%%%%%%%%%
The key point of Wolfenstein's paper \cite{w78a} 
is that  ``Coherent forward  scattering  of neutrinos must be taken 
into account when considering oscillations in matter''. 
%%At this point he thanks Zavattinni  for .... 
%%(which one?)  indicating that matter effect is impotant. 
Here Wolfenstein  used analogy with regeneration of $K_S$ from 
the $K_L$ beam (see a comment below), as well as with optics, 
without discussion of validity and applicability of the analogy. 

Matter effect is described by the refractive indices which have definite values 
for the eigenstates of the NC interactions (\ref{eq:eigen}):
\begin{equation}
n_i =  1 + \frac{2\pi N}{k^2} f_i(0).   
\label{eq:refind}
\end{equation}
Here $N$ is the effective number density of scatterers, 
$f_i(0)$ is the  amplitude of   $\nu_i$ forward scattering 
and $k$ is the neutrino momentum. 

The refractive index modifies the phase of propagating state:
$e^{i k n_i x} \nu_i$. 
The  phase difference (relevant for oscillations) equals 
\begin{equation}
k \Delta n x =  2\pi N x \frac{\Delta f(0)}{k},  
\label{eq:phased}
\end{equation}
where $\Delta n \equiv n_2 - n_1$ and $\Delta f \equiv f_2 - f_1$. 
Without explanations,  Wolfenstein presents final result:
\begin{equation}
k \Delta n = 2 G_F \sin^2\alpha ~\Sigma_k g_k N_k, ~~~k = p,~ n,~ e. 
\label{eq:kdelta}
\end{equation}

{\it 2. Refraction length and  scale of the effect.}
Refraction length $l_0$  is  the distance over which the phase 
difference (\ref{eq:phased}) equals $2\pi$: $k \Delta n l_0 = 2\pi$, 
which gives  
\begin{equation}
l_0 = \frac{2\pi}{k \Delta n} \sim \frac{1}{G_F N_A} 
\label{eq:refrlength}
\end{equation}
($N_A$ is the Avogadro number). 
For massless neutrinos, when (\ref{eq:kdelta}) is 
the only source of phase 
difference, $l_0$ equals the oscillation length: $l_0 = l_m$. 
Numerically,  $l_0 \approx  10^9$ cm is comparable with the radius 
of the Earth. So, the matter effect can be observed in experiments with 
baselines $\geq 10^8$ cm. Wolfenstein refers to  Mann and Primakoff's  paper 
\cite{mann} where detection of neutrinos in Quebec (Canada) 1000 km  
away from their source at Fermilab was proposed. 

According to (\ref{eq:refrlength}) the refraction length 
does not depend on neutrino energy. Furthermore, at low energies 
$l_{inel} \gg l_0$   and the inelastic interactions can be neglected. \\ 

{\it 3. Oscillations.} Wolfenstein  introduced the notion 
of the {\it eigenstates for propagation in matter}.  
The eigenstates (\ref{eq:eigen}), that  diagonalize the Lagrangian of NC interactions 
(\ref{eq:hamilton}),  have definite refraction indices $n_i$  
and therefore acquire definite phases.  
These states differ from $\nu_a$ and $\nu_b$ - 
the neutrino states produced in the charged current 
interactions, and this means mixing. 

Evolution of neutrino states produced in the CC interactions is  
given by 
\begin{equation}
\nu_a(x) = \cos \theta_m \nu_{1m} e^{ikn_1 x} 
       + \sin \theta_m \nu_{2m} e^{ik n_2x}.    
\label{eq:evolutiona}
\end{equation}
From here the derivation of  expression for the oscillation 
probability is straightforward: $|\langle \nu_a | \nu_a (x)\rangle|^2 = 
0.5[1 + \cos k\Delta n x]$.    
Wolfenstein considered maximal mixing
and therefore oscillations with maximal depth. 
The oscillation length equals $l_0$. \\

{\it 4. Charged current contribution.} In the footnote of \cite{w78a} Wolfenstein writes 
``I am indebted to Dr. Daniel Wyler for pointing 
out the importance of the charged-current term".   
Daniel (who was in Carnegie-Mellon with Wolfenstein before he 
moved to Rockefeller in 1977) told me the story.    
``Lincoln had just written and presented at a meeting at Fermilab a short paper 
... [on] neutrino oscillations in matter... 
I took the preprint with me over the weekend and discovered that Lincoln had forgotten 
the charged current interactions. I then called him on Monday morning to tell him. 
His paper had already been accepted by PRL and had to be retracted.'' 
%%There was some excitement [about possible joint word. 
A collaboration did not develop: 
``I myself worked other things and did not think much about neutrinos; 
also Lincoln was quite secretive about this stuff and did not want
anyone be part of it."
%Therefore he did not discuss things very much.''

%%Apparently, the corresponding part of the text 
%%was added to the paper latter. 
Wolfenstein writes in the revised and extended version of the paper  
that if one of the oscillating neutrinos is $\nu_e$, the CC scattering on electrons
%%, Fig.~\ref{fig:ccdiag},  
contributes to the phase difference.
Using the Fierz transformation this contribution 
is reduced to the NC contribution, {\it i.e.} to the elastic 
forward scattering relevant for refraction. 
This gives $k \Delta n = - G_F N_e$ (later corrected to $\sqrt{2} G_F N_e$), which in fact, is the 
standard matter potential called the Wolfenstein potential.  

%%%%%%%%%%%ffff1%%%%%%%%%%%%%%%%%%%%%%%%%%%%%%%%%%%%%%%%%%%%%%%%%%%%%%%%%%%%
%%\begin{figure}
%%\begin{center}
%%\begin{minipage}{0.33\linewidth}
%%\centerline{\includegraphics[width=1.0\linewidth]{diagr1}}
%%\end{minipage}
%%\end{center}
%%\caption[]{Diagram for scattering of the electron neutrino 
%%on electron due to the charged currents.}
%%\label{fig:ccdiag}
%%\end{figure}
%%%%%%%%%%%%%%%%%%%%%%%%%%%%%%%%%%%%%%%%%%%%%%%%%%%%%%%%%%%%%%%%%%%%%%%%%%
 
The CC contribution (i) changes the mixing angle and oscillation length 
of massless neutrinos; (ii) modifies the vacuum oscillations even 
when NC are diagonal and symmetric as in the Standard Model. \\
%%[[The eigenstates  of propagation ]]

{\it 5. Modification of oscillations of massive neutrinos.} 
%%%%%%%%%%%%%%%%%%%%%%%%%%%%%%%%%%%%%%%%%%%%%%%
For massive neutrinos another source 
of phases and phase difference exists (apart  
from coherent scattering) which is related to masses:
$$
|\nu_i (t)\rangle = e^{-i m_i^2 t/2k} |\nu_i \rangle .     
$$
This contribution is well defined in the mass basis,  
while the matter effect -- in the interaction basis. 
To accommodate both contributions Wolfenstein derived   
the differential equation \cite{w78a}: 
\begin{equation}
i\frac{d}{dt}
\left(\begin{array}{c}
\nu_1\\
\nu_2
\end{array} \right)
 =
\left( \begin{array}{cc}
\frac{m_1^2}{2k} -  G N_e \cos^2\theta  & - GN_e \sin \theta \cos \theta  \\
 - GN_e \sin \theta \cos \theta  &  \frac{m_2^2}{2k} - G N_e \sin^2\theta
\end{array} \right)
\left(\begin{array}{c}
\nu_1\\
\nu_2
\end{array} \right), 
\label{eq:mastereq}
\end{equation}
where $s \equiv \sin \theta$,  $c\equiv \sin \theta$        
and $\theta$ is the vacuum mixing angle. 
This is the evolution equation in the mass basis. 
Apparently, the corresponding part of the text 
was added to the paper latter. \\

{\it 6. The parameters of oscillations.}
%%%%%%%%%%%%%%%%%%%%%%%%%%%%%%%%%%%%%%%%%%%%%%%%%%%%%%
Mixing angle in matter $\theta_m$ relates the eigenstates 
for propagation in matter and the flavor states \cite{w78a}. Wolfenstein found  
\begin{equation}
\tan2\theta_m = \tan 2 \theta \left[1 - 
\frac{l_\nu}{l_0}  \cos^{-1}2 \theta \right]^{-1},   
\label{eq:mixangle}
\end{equation}
and the  oscillation length in matter  
\begin{equation}
l_m =  l_\nu \left[1 +  \left(\frac{l_\nu}{l_0} \right)^2          
- 2\cos2\theta~ \frac{l_\nu}{l_0} \right]^{-1/2} . 
\label{eq:osclength}
\end{equation}
The transition probability in matter with constant density equals \cite{w78a}  
\begin{equation}
P = \frac{1}{2}\sin^2 2\theta \left(\frac{l_m}{l_\nu}\right)^2 
[ 1 - \cos (2\pi x/l_m)].   
\label{eq:probosc}
\end{equation}

Three cases were noticed:

1. $l_\nu \ll l_0$ - nearly vacuum oscillations;  $l_m \approx l_\nu$ 
and $\theta_m \approx \theta$.  

2. $l_\nu \gg l_0$ - matter dominance case; $l_m \approx l_0$ and 
$\sin 2 \theta_m \rightarrow 0 $. 

3. $l_\nu \approx l_0$ - intermediate case 
$l_m \approx l_\nu$; here ``the quantitative results in matter 
are quite different from those in vacuum''. 
 
For the last case Wolfenstein gave the  
table with values of the transition probability for $l_\nu = l_0$ 
(which, in fact, is close to the resonance for small mixing). 
In particular,  for $\theta = 15^{\circ}$ and ${x}/{l_0} =0.5$ 
he obtained enhanced probability  $P = 0.492$, while  $P = 0.250$ in vacuum. 
There is no further discussion of this the most interesting case. 
%%of $l_\nu  \approx l_0$. 

Wolfenstein marked what we call now the {\it vacuum mimicking}:
``independent of the value $l_\nu /l_0$, 
as long as oscillation phase is small, $2\pi x/l_m  \ll 1$,  
the oscillation probability in the medium (\ref{eq:probosc}) 
is approximately the same as in vacuum". \\

\subsection{Follow-up}
%%%%%%%%%%%%%%%%%%%%%%%%%%%%%%%%%%%%%%%%%%%%%%%%%%%

In the paper~\cite{w79a} ``Effects of matter on neutrino oscillations" 
Wolfenstein refined the discussion and added few clarifications: 
``Oscillations of massless neutrinos are analogous 
to the phenomenon of optical {\it birefringence} in which 
case two planes of polarization are eigenvectors 
and beams with other states of polarization are 
transformed as they  pass through the crystal". 

%%Applications to the atmospheric neutrinos have been considered.  

The evolution (``master") equation was written 
in the flavor basis~\cite{w79a}:
\begin{equation}
i\frac{d}{dt}  
\left(\begin{array}{c}
\nu_e\\
\nu_\mu
\end{array} \right) 
 =
- \frac{\pi}{l_\nu}
\left( \begin{array}{cc}
\cos 2\theta  - 2(l_\nu/l_0)  &   \sin 2\theta \\
\sin 2\theta  &   - \cos 2\theta 
\end{array} \right) 
\left(\begin{array}{c}
\nu_e\\
\nu_\mu
\end{array} \right).    
\label{eq:master}
\end{equation}
Wolfenstein reiterated that in the standard case, the  CC interactions of $\nu_e$   
with electrons change  the phase of $\nu_e$ relative 
to $\nu_\mu$. This differs from the case of $\nu_\mu$  and  $\nu_\tau$. 

L. Wolfenstein submitted contribution with the same title as~\cite{w79a}
 to the proceedings of ``Neutrino-78'' in Purdue university \cite{nu78}.  
%%There was no talk.  
The adiabaticity for massless neutrinos was introduced which  I will discuss 
later in the way I have found this ``unknown" paper. \\

{\it  Applications} 
%%%%%%%%%%%%%%%%%%%%%%%%%%%%%%%%%%%%%%%%%%%%%%%%%%%%%%

1. LBL experiment for  searches for matter effects on oscillations  
was pointed out in  \cite{w78a}.

As far as solar and supernova neutrinos are concerned, 
Wolfenstein focused on the suppression of oscillations 
(in constant density media). 

2. Solar  neutrinos: he writes ``if $l_\nu$ is large, the oscillation 
should be calculated for actual vacuum path ignoring 
passage through  matter. There are no significant 
oscillations inside the Sun or in transversals 
through the earth" \cite{w78a}. [A.S.:  the adiabatic conversion is completely missed.] 

3. Supernova neutrinos \cite{w79sn}: ``Vacuum oscillations are 
effectively inhibited from occurring  because of high density''.  
The mixing
$$
\sin^2 2\theta_m \approx \sin^2 2\theta 
\left(\frac{l_0}{l_\nu}\right)^2 
$$
is very small. 

4. Atmospheric neutrinos \cite{w79a}: In massless case the survival probability 
was computed as a function of the 
zenith angle of neutrino trajectory for 
different values of $\alpha$  defined in Eq. (\ref{eq:currents}).\\

{\it Comments and remarks}
%%%%%%%%%%%%%%%%%%%%%%%%%%%%%%%%%%%%%%%%%%%%%%%%%%%%%%%%

1. Refraction of neutrinos was considered 
before Wolfenstein by R. Opher: In the paper 
``Coherent scattering of Cosmic Neutrinos", \cite{opher}
devoted to possibility to detect the relic neutrinos,  the 
expression for the refraction index $n$ was found: 
$n - 1 = \sqrt{2} G_F N/E$. 
%%[[Wolfenstein does not refer to this paper]]
The refraction index was correctly computed in~\cite{Langacker:1982ih}.

2. In the acknowledgment of \cite{w78a} Wolfenstein  
thanks  E. Zavattini for ``asking the right 
question''. What was this? Zavattini was working on 
birefringence that time.   D. Wyler writes  
``...
%I certainly remember that Lincoln referred 
%to Zavattini and mentioned his question. 
I do not remember or never knew the question. Maybe the question was whether
neutrinos could  regenerate like kaons." 
[A.S.: That would be, indeed,  the key question!]  
And he adds:  
``I would say,  his [Wolfenstein's]  main insight was that in forward direction there is a term
proportional to $G_F$ and not $(G_F) ^2$".

3. Wolfenstein discussed extreme situations but not much the 
most interesting case $l_\nu \approx l_0$. 
Surprisingly (for a physicist), even the pole 
in $\tan 2 \theta_m$  dependence on $l_\nu /l_0$  
(see (\ref{eq:mixangle})) was overlooked or ignored.

%%%%%%%%%%%ffff2%%%%%%%%%%%%%%%%%%%%%%%%%%%%%%%%%%%%%%%%%%%%%%%%%%%%%%%%%%%%
\begin{figure}
\begin{center}
\begin{minipage}{0.33\linewidth}
\centerline{\includegraphics[width=1.2\linewidth]{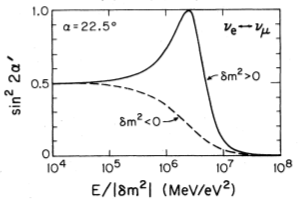}}
\end{minipage}
\, \, \, \, \, 
\begin{minipage}{0.33\linewidth}
\centerline{\includegraphics[width=1.2\linewidth]{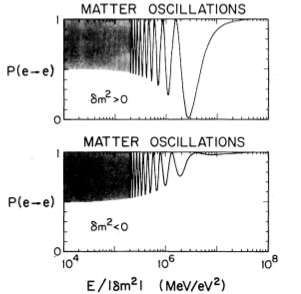}}
\end{minipage}
\end{center}
\caption[]{Dependences of the amplitude of oscillations in matter 
$\sin^2 2\alpha'$  (left),  and  the survival probability 
on  energy ($E/\delta m^2$) (right). 
Value of the vacuum mixing parameter $\sin^2 2\alpha = 0.5$ (from \cite{barger80}).   
}
\label{fig:peak}
\end{figure}
%%%%%%%%%%%%%%%%%%%%%%%%%%%%%%%%%

\subsection{Matter effects on three-neutrino oscillations. Condition of maximal mixing}
%%%%%%%%%%%%%%%%%%%%%%%%%%%%%%%%%%%%%%%%%%%%%%%%%% 

V. D. Barger, K. Whisnant, S. Pakvasa and 
R. J. N. Phillips \cite{barger80} considered the standard case: 
vacuum mixing, no FCNC,  the CC scattering on electrons, 
constant density.  
(i) Correct expression for the refraction length is given:
$l_0 = 2\pi/(\sqrt{2} G_F N_e)$.  (ii) Expressions for the oscillation probabilities 
in terms of the level splitting $\Delta M_{ij}$ were computed.     
Explicit (rather lengthy) analytic formulas for $\Delta M_{ij}$ 
were presented in the $3\nu$ case.  
(iii) A number of statements, which we use now,  appeared for the first time. 
In particular, `` Matter effect resolves the vacuum oscillation ambiguity 
in sign of $\Delta M_{ij}$'',  the  matter effect is different for neutrinos 
and antineutrinos. \\  

{\it Enhancement of oscillations} 
%%%%%%%%%%%%%%%%%%%%%%%%%%%%%%%%%%%%%%%%%%
(already noted by Wolfenstein in \cite{w78a}) was  
stated \footnote{I copy the text as it appears in \cite{barger80}. }. 
``There is always some energy, where  
%%\begin{equation}
$l_\nu / l_0  = \cos 2\theta$, 
%\label{eq:rescond}
%\end{equation}
and hence  $\theta_m = 45^{\circ}$ for either  $\nu$ or $\bar{\nu}$ 
depending on the sign of $\Delta m^2$. 
Hence there is always some energy where  $\nu$ or $\bar{\nu}$
matter mixing is maximal [AS.: see Fig. \ref{fig:peak}, left]. 
At this energy the survival probability 
vanishes at a distance 
$$
L = \frac{1}{2} l_0 \cot 2 \theta."  
$$
Numerous plots with dependences of the oscillation probabilities in matter on energy 
were presented for $2\nu$ (see e.g. Fig. \ref{fig:peak}, right) and $3\nu$ mixings.

%%[A.S.:That is, oscillations proceed with maximak depth]
%%Eq. (\ref{eq:rescond}) 
%%the maximal mixing condition is nothing but the resonance condition  
%%introduced later in our paper \cite{ms1}. 

%%%%%%%%%%%%%%%%%%%%%%%%%%%%%%%%%%%%%%%%%%%%%%%%%%%%%%%%%
\section{Mikheyev and Smirnov mechanism, 1984  - 1985}
%%%%%%%%%%%%%%%%%%%%%%%%%%%%%%%%%%%%%%%%%%%%%%%%

In 80ies both Mikheyev and myself 
worked in the  Department of Leptons of High Energies and Neutrino 
Astrophysics, ``OLVENA",  of INR of the USSR Academy of Sciences,   
led by  G. T. Zatsepin.  The department was mostly an experimental one,   
dealing with solar neutrino spectroscopy (Gallium, Chlorine, 
Li experiments), supernova neutrinos (Artemovsk, Baksan, LSD), 
cosmic rays (Pamir) and cosmic neutrinos. 
A part of the department headed  by 
A.E. Chudakov  was running experiments at Baksan Neutrino 
Telescope, on cosmic rays, atmospheric neutrinos, {\it etc.}  

Stanislav Mikheyev (1940 - 2011) was an experimentalist working at the Baksan 
telescope.   He was responsible for analysis of the atmospheric neutrino data and  
searches for oscillations (actually,  the first searches). 
Later he joined MACRO, K2K,  
Baikal neutrino telescope collaborations. 
I belonged to a small group of theorists, and topics of my research included 
cosmic neutrinos of high energies (papers with V. Berezinsky),  neutrino decay,  GUT's, {\it etc.}
%% I was affashinated by GUT.

By the way, at the Moriond 1980 on request of the collaboration I presented the first bound 
on oscillations of the atmospheric neutrinos obtained at the Baksan telescope.

%%%%%%%%%%%ffff3%%%%%%%%%%%%%%%%%%%%%%%%%%%%%%%%%%%%%%%%%%%%%%%%%%%%%%%%%
\begin{figure}
\begin{center}
\begin{minipage}{0.33\linewidth}
\centerline{\includegraphics[width=1.0\linewidth]{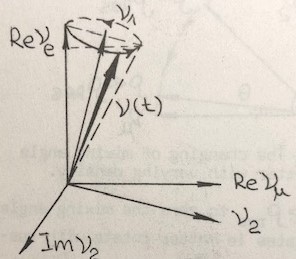}}
\end{minipage}
\end{center}
\caption[]{Graphic representation of neutrino oscillations.}
\label{fig:graph-osc}
\end{figure}
%%%%%%%%%%%%%%%%%%%%%%%%%%%%%%%%%%%%%%%%%%%%%%%%%%%%%%%%%%%%%%%%%%%%%%%%%%

In the beginning of 80ies my interest in neutrino oscillations was triggered by Bilenky and Pontecorvo's 
review~\cite{Bilenky:1978nj}.   
I had constructed a  geometrical representation of oscillations,  Fig.~\ref{fig:graph-osc},  
which later played a crucial role for our understanding of   
the MSW effect.  The representation was for neutrino states (amplitudes) and 
not probabilities,  which we use now. 
The neutrino state produced as $\nu_e$ evolves as 
$
\nu(t) = \cos \theta \nu_1  + \sin \theta e^{i\phi} \nu_2. 
$ 
Therefore in the basis formed by $(\nu_1,~ \nu_{2R}, ~ \nu_{2I})$
it can be represented as unit vector with coordinates 
$(\cos \theta, ~\sin \theta \cos\phi,~ \sin \theta \sin\phi) $.   
This basis is turned with respect to the flavor basis formed 
by $(\nu_e, ~\nu_{\mu R}, ~\nu_{\mu I})$ by 
the mixing angle $\theta$. With change of $\phi(t)$ the neutrino vector 
precesses around  $\nu_1$. 
The amplitude of probability to find $\nu_e$ in 
$\nu(t)$ is approximately equal to the projection of $\nu(t)$ onto 
the axis $\nu_e$. It equals exactly when $\nu(t)$ is in the real plane 
$(\nu_e, ~\nu_{\mu R})$ or $(\nu_1, ~\nu_{2 R})$. 
Indeed, the projection equals $(\vec{\nu}(t) \cdot \vec{\nu}_e ) 
= c^2 + s^2 \cos \phi$,  whereas the exact expression for the 
amplitude has in addition the imaginary component:  
$A_e = c^2 + s^2 \cos \phi + i s^2 \sin \phi$. The projection 
would reproduce the amplitude exactly, if one adds  the imaginary 
components to the flavor basis in the direction $\nu_{2 R})$:  
$\vec{\nu}_e = (c, s, i s)$, $\vec{\nu}_\mu = (-s, c, i c)$. 
This picture allowed me to obtain 
qualitative and in many cases - quantitative results. 
I gave seminars on that in INR.

%%I had wrong idea to improve sensitivity to oscillations: 
%%instead of long distance, use long  time of neutrino emission 
%%(use long lived isotopes). 
%%A Linde discussed ... 

The starting point of the ``M-S collaboration'' was sometime in  February - March of 1984 when  
Stas Mikheyev asked me if I  know the Wolfenstein's paper.  
``Is it correct? Should matter effects be taken into account 
in the oscillation analysis of the atmospheric neutrinos?''
I did not know Wolfenstein's paper.  
Stas gave me the reference and I started to read it.

\subsection{Resonance} 
%%%%%%%%%%%%%%%%%%%%%%%%%%%%%%%%%%%%%%%%%%%%%%%%%%%%%%%%%%%%%%%%%%%%%%%

\begin{figure}
\begin{center}
\begin{minipage}{0.33\linewidth}
\centerline{\includegraphics[width=1.1\linewidth]{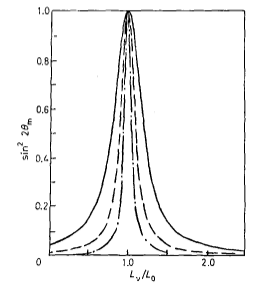}}
\end{minipage}
\end{center}
\caption[]{Resonance. Dependence of the mixing parameter in matter 
$\sin^2 2\theta_m$ on $l_\nu/l_0 \propto E N$ for different values of 
the vacuum mixing parameter: $\sin^2 2\theta = 0.04,~0.01,~2.5 \cdot 10^{-3}$.
From~\cite{ms1,ms2}.}
\label{fig:resonance}
\end{figure}

One of the first things I did was to 
draw  the mixing parameter in matter  $\sin^2 2\theta_m$ 
%%(which determines the depth of oscillations) 
as function of  $l_\nu/l_0$,  
%%written in the form  
%%mixing angle obtained from (\ref{}):  
\begin{equation}
\sin^2 2\theta_m = \frac{\sin^2 2\theta}
{1 - 2 (l_\nu /l_0)\cos 2\theta +  (l_\nu/l_0)^2 },   
\label{eq:sinangm}
\end{equation} 
for different values of the vacuum 
mixing angle. The result (Fig.~\ref{fig:resonance}) was astonishing! 
For small values of $\sin^2 2\theta$
dependence of $\sin^2 2\theta_m$ on $l_\nu/l_0$ 
has a resonance form. At the condition 
\begin{equation}
\frac{l_\nu}{l_0}  = \cos 2\theta, 
\label{eq:rescond1}
\end{equation}
which we called the {\it resonance  condition}
the depth of oscillations reaches maximum: 
$\sin^2 2\theta_m =  1$ \cite{ms1,ms2}. 
The condition (\ref{eq:rescond1}) coincides with condition of maximal mixing  
in \cite{barger80}. 

%%%%%%%%%%%ffff4%%%%%%%%%%%%%%%%%%%%%%%%%%%%%%%%%%%%%%%%%%%%%%%%%%%%%%%%%%%%
%%%%%%%%%%%%%%%%%%%%%%%%%%%%%%%%%%%%%%%%%%%%%%%%%%%%%%%%%%%%%%%%%%%%%%%%%%

Immediate question was about the  nature of the peak in $\sin^2 2\theta_m$.  
Is it accidental or has certain physical meaning? 
We started to explore different aspects of the resonance being 
very much surprised that nobody noticed this before. 
For small vacuum mixing the resonance condition becomes 
\begin{equation}
 l_\nu \approx l_0 
\label{eq:rescondsmall}
\end{equation}
- the vacuum oscillation length approximately 
equals the refraction length. 
That is, the eigenfrequency of the system, $1/l_\nu$,  equals 
the eigenfrequency of ``external'' medium $1/l_0$.   
This sounds as a resonance in usual sense. 
The width of the resonance at the half-height equals
\begin{equation}
\Delta (l_\nu /l_0) = 
(l_\nu /l_0)_{res} \tan 2\theta = \sin2 \theta,  
\label{eq:reswidth}
\end{equation}
{\it i.e.}, the smaller vacuum mixing  (the weaker binding in the system),  
the narrower resonance - another signature  of real resonance. 
For large vacuum mixing (strong binding)
the resonance shifts (deviates from (\ref{eq:rescondsmall})) as expected.  
%%yet another confirmation.   
Later we introduced the resonance factor $R \equiv \sin^2 2\theta_m / \sin^2 2\theta$,  
which reproduces another feature of true resonance: 
the smaller the  mixing $\theta$, the higher the peak. 
The oscillation length in resonance becomes maximal:
$l_m^R = l_\nu /\sin 2 \theta$. 
Two different manifestations of the resonance  
were identified:   
(i) resonance enhancement of oscillations 
in constant density for continuous neutrino spectrum; 
(ii) adiabatic conversion in varying density and for monoenergetic neutrinos.   

%%%%%%%%%%%%%%%ffff %%%%%%%%%%%%%%%%%%%%%%%%%%%%%%%%%%%%%%%%%

\subsection{Resonance enhancement of oscillations}
%%%%%%%%%%%%%%%%%%%%%%%%%%%%%%%%%%%%%%%%%%%%%%%%%%%%%%%%%%

At the resonance energy determined by the condition 
(\ref{eq:rescond1}) \cite{ms1,ms2} 
\begin{equation}
E_R = a \frac{\Delta m^2 \cos2\theta}{\rho}, ~~~~
a \equiv m_N /2\sqrt{2} G_F Y_e    
\label{eq:resener}
\end{equation}
oscillations proceed with maximal depth. In the energy interval 
determined by the width of the resonance, 
$\Delta E_R = E_R \tan 2\theta$,  oscillations are enhanced. 
%%This reproduces findings in \cite{barger80}.

%%[[remove?]] One comment is in order. 
Recall that already Wolfenstein found an enhancement of the transition probability 
due to matter effect for certain values of parameters (which correspond to $l_\nu \sim l_0$), but 
left this without discussion. 
Barger {\it et al.},  wrote the condition for maximal 
mixing and showed enhancement of oscillations, but the resonance nature was not uncovered.  
I read the paper \cite{barger80} already after 
we had realized the existence of the resonance. I did not find the term resonance in~\cite{barger80};  
and dependence of the mixing in matter on  
energy shown for large vacuum mixing (Fig.~\ref{fig:peak})  
looked in the log scale as a peak at the end of shoulder. We cited  
the paper~\cite{barger80}  in connection to a possibility of  
enhancement of oscillations in matter \cite{ms1}, but did not 
comment on the maximal mixing condition. 
In this connection we received ``clarification'' letter from Sandip Pakvasa.

%%[[to conclusion]] MS realized the resonance nature of the matter 
%%effect, introduced notion of resonance, studied nature 
%%and properties of the resonance.  

%% In particular, 
%%- The smaller the vacuum mixing (strength of coupling)
%% the narrower resonance, 
%%- Shown that it has the same features 
%%  as resonances in other systems.  
%%- Explored possible manifestations of the resonance. 

%%%%%%%%%%%%%%%%%%%%%%%%%%%%%%%%%%%%%%%%%%%%%%%%%%%%%%%%%%
\subsection{Varying density}
%%%%%%%%%%%%%%%%%%%%%%%%%%%%%%%%%%%%%%%%%%%%%%%%%%%%

For neutrinos with a given energy $E$ propagating in varying density medium   
significant enhancement of oscillations (transition) occurs 
in the resonance layer with density 
\begin{equation}
\rho_R = a\frac{\Delta m^2 \cos2 \theta}{E}
\label{eq:resdensity}
\end{equation}
(again determined by (\ref{eq:rescond1})) and width
\begin{equation}
\Delta \rho_R = \rho_R \tan 2\theta.   
\label{eq:deltaro}
\end{equation}
Spatial width of the resonance layer equals 
$$
r_R = \left(\frac{d\rho_R}{dr}\right)^{-1} \Delta \rho_R.
$$
Resonance enhancement of transitions is significant  
if the resonance layer is sufficiently thick: 
\begin{equation}
r_R  >  l_m^R  = l_\nu/\sin2\theta, 
\label{eq:adcond1}
\end{equation}
that is, the width of resonance layer is  larger 
than the  oscillation length in resonance. \\ 

%%We understood that relations for constant density may not be  
%%applicable for varying density, but concepts 
%%of resonance layer, its density and width are 
%%useful for qualitative analysis.  
%%Furthermore, condition for strong transformation 
%%(\ref{eq:adcond1}) reproduces correctly the  adiabaticity condition.  \\

\begin{figure}
\begin{center}
\begin{minipage}{0.33\linewidth}
\centerline{\includegraphics[width=1.9\linewidth]{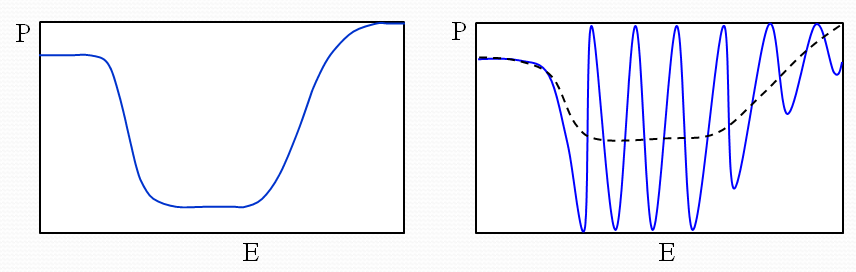}}
\end{minipage}
\end{center}
\caption[]{Dependence of the survival probability on energy for 
adiabatic conversion (left), and in the ``slab model'' (right).}
\label{fig:ad-vs-slab}
\end{figure}

At this point we encountered a puzzle. 
In medium with varying density  (like the Sun) 
both the resonance condition and condition 
for strong transformation (\ref{eq:adcond1})
are satisfied in wide energy range, so one would expect 
strong transitions in this range. 
Our first guess for the survival probability is shown in  Fig.~\ref{fig:ad-vs-slab}, left. 
The left edge of the ``bath" 
is given by the resonance energy which corresponds to   
maximal density in the medium. At lower energies  
there is no resonance  and therefore no strong transitions.
At high energies since  $l^m_R \propto E$  and  $r_R = const$ (the Sun), 
the condition $r_R > l^m_R$  is broken: the resonance layer is too narrow 
for developing strong transition.
What happens in the intermediate energy range?
%
%%%%%%%%%%%ffff5%%%%%%%%%%%%%%%%%%%%%%%%%%%%%%%%%%%%%%%%%%%%%%%%%%%%%%%%%%%%
%%%%%%%%%%%%%%%%%%%%%%%%%%%%%%%%%%%%%%%%%%%%%%%%%%%%%%%%%%%%%%%%%%%%%%%%%%
%
It was a confusion in the  spirit of the later  
``slab" model by Rosen and Gelb \cite{Rosen:1986jy}.  
If oscillations with large amplitude occur in the resonance 
layer, why the phase of oscillations at the end of the layer 
is always close to  $\pi$ and does not change with energy? 
Why there is no oscillatory picture as in 
Fig.~\ref{fig:ad-vs-slab}, right?
Numerical computations confirmed the result of the left panel.

\subsection{Numerical solution}
%%%%%%%%%%%%%%%%%%%%%%%%%%%%%%%%%%%%%%%%%%%%%%%%%%%

We introduced the bi-linear forms of the wave functions   
\begin{equation}
P \equiv \nu_e^* \nu_e, ~~~~R + iI = \nu_\mu^* \nu_e    
\label{eq:nuvector}
\end{equation}
which are,  in fact, the elements of density matrix, or   
equivalently, components  of the neutrino polarization vector in the flavor space.  
Then using the  Wolfenstein's evolution equation (\ref{eq:master}) 
for the wave functions we derived  the 
system of equations for  $P,~ R,~ I$:  
\begin{equation}
\frac{dP}{dt} = - 2M I, ~~~~~
\frac{dI}{dt} = - m R + M(2P -1), ~~~~~
\frac{dR}{dr} = m I. 
\label{eq:threeeq}
\end{equation}
Here 
\begin{equation} 
2M \equiv   \frac{2\pi}{l_\nu} \sin2 \theta, ~~~~
m \equiv  \frac{2\pi}{l_\nu} 
\left(\cos 2\theta - \frac{l_\nu}{l_0} \right).     
\label{eq:pridef}
\end{equation}
$P$ is the $\nu_e$ survival probability. 
If $\nu_e$  is produced, the initial conditions read 
$$
P(0) = 1,~~  I(0) = R(0) = 0. 
$$
We thanked N. Sosnin, my classmate in MSU,  
for indicating Runge-Kutta method to solve the equation. 
We overlooked that Eqs. (\ref{eq:threeeq}) can be written in the vector product form. 

\subsection{Towards the adiabatic solution}
%%%%%%%%%%%%%%%%%%%%%%%%%%%%%%%%%%%%%%%%%%%%%%

\begin{figure}
\begin{center}
\begin{minipage}{0.33\linewidth}
\centerline{\includegraphics[width=1.2\linewidth]{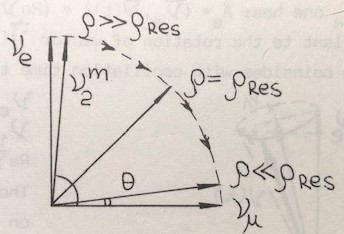}}
\end{minipage}
\hskip 1cm
\begin{minipage}{0.33\linewidth}
\centerline{\includegraphics[width=1.0\linewidth]{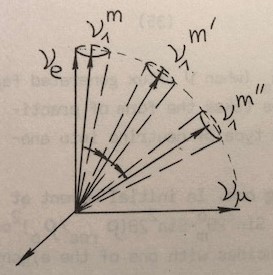}}
\end{minipage}
\end{center}
\caption[]{
Graphic representation of the adiabatic conversion. 
{\it Left:} Rotation  of the eigenstate vector $\nu_{2m}$ with decrease of density. This shows neutrino  
evolution in the case of non-oscillatory transition since for large initial density $\nu(t) \approx \nu_{2m}$. 
{\it Right:} Shown is evolution of the system when neutrino produced as $\nu_e$ propagates toward large densities. 
From~\cite{tignes}
and WIN-85 slides \cite{win}.}
\label{fig:grap-adc}
\end{figure}

To understand the results of the numerical computations we used graphic representation. 
With changing density the mixing in matter changes,  
Fig. \ref{fig:grap-adc}, left. The mixing angle determines the direction 
of the cone axis on which the neutrino vector precesses. 
If density (and therefore the mixing angle in matter)
changes slowly, the system (neutrino vector) has 
{\it time to adjust itself} to these changes, so the precession cone turnes   
together with the axis $\vec{\nu}_1$ and the cone angle 
does not change (Fig.~\ref{fig:grap-adc}, right). \\
%%This allowed to explain numerical results.\\
 
%
%%%%%%%%%%%ffff6%%%%%%%%%%%%%%%%%%%%%%%%%%%%%%%%%%%%%%%%%%%%%%%%%%%%%%%%%
%%%%%%%%%%%%%%%%%%%%%%%%%%%%%%%%%%%%%%%%%%%%%%%%%%%%%%%%%%%%%%%%%%%%%%%%%%

%%Fig.~\ref{fig:grap-adc} From Moriond 86 \cite{tignes}, 
%%and similar figure was shown at WIN-85 in Savonnlina \cite{win}. 

%%%%%%%%%%%%%%%%%%%%%%%%%%%%%%%%%%%%%%%%%%%%%%%% 

{\it Wolfenstein's letter and adiabaticity.} 
%%%%%%%%%%%%%%%%%%%%%%%%%%%%%%%%%%%%%%%%%%%%%
We sent to Wolfenstein a preliminary  version of our paper. 
%%He had replied few months later. 
In short reply letter 
(unfortunately lost) he said essentially that
it should be no strong transitions inside the Sun 
due to adiabaticity and gave reference to the proceedings \cite{nu78}.  
We could not find these proceedings,
%%~\footnote{Even now one can not find this paper on Spires also 
%%in the list of papers published in proceedings since that was submitted 
%%paper and not the talk.} 
but cited his contribution and thanked him ``for a remark concerning adiabatic 
regime of neutrino propagation'' \cite{ms1}. 
We started to call the effect of {\it adjustment} 
of the system to the density change the adiabatic 
transition and the condition of strong transition, 
(\ref{eq:adcond1}) -- the adiabatic condition.    
Our reply was that  it is due to the adiabaticity 
that a strong transformation can occur. We introduced this terminology 
in  proofs of the papers \cite{ms1,ms2}. \\

%%%%%%%%%%%%%%%%%%%%%%%%%%%%%%%%%%%%%%%%%%%%%%%%%%

Wolfenstein's reply probably explains why he did 
not proceed with further developments of his ideas. 
Later, Bruno Pontecorvo told me that he had a discussion 
with Wolfenstein 
%%(where, when? - I did not asked) 
and they concluded that, 
it seems, there is no practical outcome of oscillations in matter. 
One can guess why Wolfenstein thought that adiabaticity prevents strong transitions: 
The adiabaticity ensures that result of transitions depends on the initial and final 
conditions only and does not depend on what happens in between. 
If the initial density is large and the final (vacuum) mixing is small, 
then both in the initial and in final states the mixing is strongly suppressed. 
Apparently,  Wolfenstein  missed that although the mixing 
is suppressed in the initial and final states, 
these states are different: in the initial state 
$\nu_e \approx \nu_{2m}$, while in the final state 
$\nu_e \approx \nu_{1m}$ (crossing of the resonance) 
and the angle changes from $\pi/2$ to $\approx 0$. 
Maybe this guess is wrong (see Sec. 3.9). 

We generalized our adiabaticity condition as 
\begin{equation}
\left(\frac{d\rho}{dr}\right)^{-1} \rho 
> \frac{l_\nu}{\tan^2 2\theta},   
\label{eq:adiabgen}
\end{equation}
which reduces to  (\ref{eq:adcond1}) in resonance.  
The adiabaticity parameter was introduced 
\begin{equation}
\kappa_R \equiv \frac{r_R}{l_m^R} = 
%\left(\frac{d\rho}{dr}\right)^{-1} \rho \frac{\tan^2 2\theta}{l_\nu}  = 
\rho \left(\frac{d\rho}{dr}\right)^{-1} 
\frac{\sin^2 2\theta}{l_\nu \cos 2\theta},   
\label{eq:adiabpar}
\end{equation} 
so that the adiabaticity condition becomes $k_R  > 1$. 
%%[[remove?]] The parameter can be rewritten in the following form 
%%$$
%%\kappa_R = \left(\frac{d\rho}{dr}\right)^{-1} \rho 
%%\frac{\tan^2 2\theta}{l_\nu}  = \left(\frac{d\rho}{dr}\right)^{-1} \rho 
%%\frac{\sin^2 2\theta}{l_0 \cos 2\theta}.    
%%$$
%%which means that  $\kappa_R \sim 
%%\left(\frac{d\rho}{dr}\right)^{-1} \rho^2$. \\

\subsection{Adiabatic conversion}
%%%%%%%%%%%%%%%%%%%%%%%%%%%%%%%%%%%%%%%%%%%%
Suppose~\cite{ms1,ms2}  neutrinos are produced at $\rho_{max}$ and pass through the medium  
with decreasing density $\rho_{min} \ll \rho_R (E) \ll \rho_{max}$, 
($\rho_{min} \sim 0$).  Then the initial mixing angle equals  
$\theta_m \approx \pi/2$, and therefore the initial neutrino state is 
$$
\nu_{initial} = \nu_e \approx \nu_{2m}(\rho_{max}),   
$$
that is, $\nu_e$ nearly coincides with the eigenstate 
$\nu_{2m}$ (see Fig.~\ref{fig:grap-adc}, left). Since $\nu_{2m}$ is the eigenstate in matter, 
in the course of adiabatic propagation, 
\begin{equation}
\nu (\rho) \approx   \nu_{2m}(\rho) \rightarrow \nu_{2m}(\rho_{min}).   
%%=  \nu_{2m}(0) = \nu_2  
\label{eq:adiabev}
\end{equation}        
In final state  $\rho_{min} = 0$,  so that  $\theta_m  = \theta$,  and therefore 
$\nu_{final} \approx \nu_{2m}(0) = \nu_2$.  
The amplitude to find $\nu_e$ in the final state equals 
$$
\langle \nu_e | \nu_{final} \rangle = 
\langle \nu_e | \nu_2 \rangle   = \sin \theta.    
$$
Therefore the survival probability is \cite{ms1,ms2} 
\begin{equation}
P = \sin^2 \theta
\label{eq:nonoscil}
\end{equation}
which is one of the main results of the papers~\cite{ms1,ms2}. It can be seen from graphic picture 
of Fig.~\ref{fig:grap-adc}, left, keeping in mind that the cone angle is very small in this case.  
%%Graphic representation of this transition is shown in 
%%Fig.~\ref{fig:grap-adc}, right.  
%%In the course of propagation mixing changes by 
%%$\sim \pi/2$ (for small vacuum mixing). 
 
For matter profile with decreasing density 
typical dependence of the suppression  factor, {\it i.e.}  
the survival probability $P$ (averaged over oscillations),  on the energy   
has the form of suppression ``bath",  Fig.~\ref{fig:ad-vs-slab}, left. 
At low energies it is given by the  
averaged vacuum oscillations with $P = 1- 0.5 \sin^2 2\theta$. 
At higher energies the non-oscillatory adiabatic conversion
gives $P = \sin^2 \theta$.  
%%The transition between  these two results occurs at the resonance energy which 
%%corresponds to the highest density in the layer.  
%%The shape of $P$ in the transition region 
%%is similar to the resonance curve. 
At even higher energies the non-adiabatic
conversion occurs and the survival probability increases with $E$  
approaching 1.\\ 

%%%%%%%%%%%ffff7%%%%%%%%%%%%%%%%%%%%%%%%%%%%%%%%%%%%%%%%%%%%%%%%%%%%%%%%%%%%
%%\begin{figure}
%%\begin{center}
%%\begin{minipage}{0.33\linewidth}
%%\centerline{\includegraphics[width=1.3\linewidth]{msa6.png}}
%%\end{minipage}
%%\end{center}
%%\caption[]{Dependence of the survival probability on energy. From \cite{ms1}.}
%%\label{fig:profile}
%%\end{figure}
%%%%%%%%%%%%%%%%%%%%%%%%%%%%%%%%%%%%%%%%%%%%%%%%%%%%%%%%%%%%%%%%%%%%%%%%%%

%%\subsection{}
%%%%%%%%%%%%%%%%%%%%%%%%%%%%%%%%%%%%%%%%%%%%%%%%%%%%%
%%What, when and how things were published.
%%In contrast, to Wolfenstein I can explain our case. 

A few words about publications. 
The first paper entitled 
``Resonance Amplification of Oscillations in Matter 
and Spectroscopy of Solar Neutrinos"
\cite{ms1},  had been submitted to Phys. Lett. B 
in 1984 and it was rejected with standard motivation of  
no reason for quick publication.  
The updated version has been sent to Yadernaya 
Fizika - Soviet Journal of Nuclear physics.  
In spring of 1985 the paper was almost rejected also from Yad. Fiz. 
We heard about skeptical opinion of  Bruno Pontecorvo.  
Later Bruno told me that he did not see the paper but 
one of his colleagues did and said ``rubbish". It was a general skepticism:  
``something should be wrong, somebody will eventually find this''.   

%%Bruno continued  ``he (colleague) did not understood anything!''  

With his usual wisdom, G.T. Zatsepin told me  
``if the paper is wrong, people will probably forget it, 
if correct - it is very important.'' 
G.T.  brought the  paper to Italy and asked C. Castagnoli 
(a collaborator in the LSD experiment) 
to consider it for publication in Nuovo Cimento 
where Castagnoli was an editor.     
 The paper  (slightly  modified) has been soon accepted to Nuovo Cim.
Suddenly it was also accepted by Yad. Fiz.  (editor I. Yu. Kobzarev).  
We made some corrections at the proofs. The content of the two papers is rather similar, 
although there are differences, e.g., in Nuovo Cimento \cite{ms2} we commented 
on effects in the $3\nu$  mixing case.

\subsection{WIN-1985}
%%%%%%%%%%%%%%%%%%%%%%%%%%%%%%%%%%%%%%%%%%%%

A. Pomansky recommended Matts Ross (the chairman) to invite me and Mikheyev to the  
WIN 1985 conference in  Savonlinna (Finland, June 16 - 22, 1985), but no talk was arranged. 
Upon arrival I described our results to Serguey Petcov whom 
I knew before (Serguey gave the plenary review talk on  Massive Neutrinos).  
%%Neutrino Oscillations and Neutrinoless Double 
%%  Beta Decay'' 
I asked to give a talk  the organizers of parallel sessions 
Gianni Conforto (``Neutrino oscillations") and 
Cecilia Jarlskog (``Status of electroweak theory"). Both said that there was no time, 
but eventually Gianni has found about 10 min at the end of his session for my presentation. 
My slides contained the description of the resonance, resonance enhancement of 
oscillations, adiabatic conversion (mostly using graphic  
representation), and applications to the Sun (suppression pits). 
As far as I remember, about 20 people  were in the room. 
S. Petcov and  N. Cabibbo (who gave the summary  talk) 
were not present. Some participants of the workshop 
made copies of the transparencies to which I had attached a part of the text of our paper. 
That was referred in many papers later as \cite{win}.   

During excursion N. Cabibbo told me that Serguey Petcov 
had described to him the results of our paper 
and he would like to include them in the summary. 
He said ``I think the effects can be understood as the level 
crossing processes'', and he showed me the drawing similar to that  
in Fig.~\ref{fig:levelcr}, but without the $\bar{\nu}_e$ line. 
``Do you agree?'' I replied immediately 
``Yes",  because I learned about this representation before.  
In the spring of 1985 after my seminar in the theoretical 
division of INR Valery Rubakov told me that our neutrino transformations resemble  
the catalysis of proton decay when monopole propagates 
near nucleon. This has an interpretation as the level crossing phenomenon".  
%%(Superheavy Magnetic Monopoles and Proton Decay, 1981).  

Petcov devoted about one third of his review to the resonance oscillations.  
Unfortunately, I  missed  Cabibbo's talk: we (with Mikheyev) left Savonlinna the day before. 
Serguey wrote to me that Cabibbo had spent a significant part of his summary    
explaining (in his own way!) our results.  
He used the levels crossing diagram, and furthermore, a     
graphic representation of the effect which differs from ours.
%% based on analogy with the electon  spin precession  
%%in rotating magnetic field. 
These talks were important contributions to the  acceptance 
of  the idea of the resonance conversion. 
Ray Davis was another key contributor: He was visiting INR in 1985. 
I gave him copies of our paper and transparencies. He took them to the US and showed to his colleagues.

%%the important contributions of S. T. Petcov,
%  R.  Davis Jr.,  N.  Cabbibo, and  L.  B. Okun'
%  to  the acceptance of the idea of resonance oscillations.'

\subsection{Theory of adiabatic conversion}
%%%%%%%%%%%%%%%%%%%%%%%%%%%%%%%%%%%%%%%%%%%%%%%%%%%%

In spring - summer 1985  we achieved  complete understanding 
of the  adiabatic conversion. 
We have written the paper ``Neutrino oscillations 
in a variable density medium and  $\nu$ - 
burst due to the gravitational collapse  of stars'' \cite{msjetp}, 
which  I like most! 

The way we got the adiabatic solution differs from the later 
derivations.  Actually,  it gives another insight onto the adiabatic 
approximation. From equations for $P, R, I$ (\ref{eq:threeeq}) 
we derived single equation of the third order for $P$ excluding $R$ 
and $I$:  
\begin{equation}
M \frac{d^3 P}{dt^3} 
- \frac{dM}{dt}\frac{d^2 P}{dt^2}
+ M \left(M^2 + 4 \bar{M}^2\right)\frac{d P}{dt} 
- 2 \bar{M}^2\frac{dM}{dt} (2P - 1) = 0, 
\label{eq:thirdpow}
\end{equation}
where $M$ and $\bar{M}$ are defined in (\ref{eq:pridef}). 
The initial conditions in the case of $\nu_e$ production read  
\begin{equation}
P(0) = 1, ~~~ \frac{dP(0)}{dt} = 0, 
~~~\frac{dP^2(0)}{dt^2} = -2\bar{M}^2.  
\label{eq:initcond}
\end{equation}
The adiabaticity means that one  can neglect the highest derivatives, 
namely, the third and the second ones.  Integration of the equation with only two last terms of 
Eq. (\ref{eq:thirdpow}) is simple. 

%%Then Eq. (\ref{eq:thirdpow}) reduces to   
%%\begin{equation}
%%\frac{d P}{dt} = \frac{2 \bar{M}^2}{\left(M^2 + 4 \bar{M}^2\right)}
%%      \frac{dM}{M dt} (2P - 1), 
%%\label{eq:thirdpow}
%%\end{equation}
%%and its integration is simple. 

\begin{figure}
\begin{center}
\begin{minipage}{0.33\linewidth}
\centerline{\includegraphics[width=1.7\linewidth]{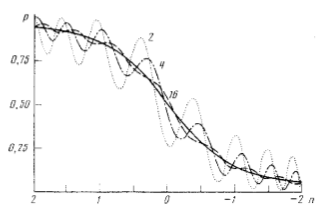}}
\end{minipage}
\end{center}
\caption[]{The MSW effect. Survival probability as function of $n$ for different 
values of $n_0$ in the production point (numbers at the curves). 
In the resonance, $n = 0$, the depth of oscillation is not maximal and decreases with increasing $n_0$. 
From~\cite{msjetp}.}
\label{fig:adconversion}
\end{figure}

Instead of the distance in space we used the distance  from the resonance in the density scale 
measured  in the units of width of the resonance layer:  
\begin{equation}
n \equiv \frac{\rho - \rho_R}{\Delta \rho_R}. 
\label{eq:nscale}
\end{equation}
In terms of $n$ the solution for average probability is \cite{msjetp}  
\begin{equation}
P(n, n_0) = \frac{1}{2}
\left[1 + \frac{n_0}{\sqrt{n_0^2 + 1}} 
\frac{n}{\sqrt{n^2 + 1}} \right].
\label{eq:adsolution}
\end{equation}
This form underlines the {\it universal} character of the adiabatic  solution 
which does not depend on density distribution, 
$n = n(r)$,  but is determined exclusively 
by the  initial and final  densities.  
Also the role of  the resonance and resonance layer can be seen  here explicitly. 
%%(I was so much exited about the resonance that used it everywhere  
%%were possible.) 
In Fig.~\ref{fig:adconversion} the survival probability is shown as function 
of $n$ for different values of $n_0$. 

%%%%%%%%%%%ffff8%%%%%%%%%%%%%%%%%%%%%%%%%%%%%%%%%%%%%%%%%%%%%%%%%%%%%%%%%%%%
%%%%%%%%%%%%%%%%%%%%%%%%%%%%%%%%%%%%%%%%%%%%%%%%%%%%%%%%%%%%%%%%%%%%%%%%%%

With increase of the initial density $n_0$ the amplitude 
of oscillations decreases and  
$P$ converges to the asymptotic non-oscillatory form
\begin{equation}
P(n, n_0 \rightarrow - \infty) = \frac{1}{2}
\left[1 + \frac{n}{\sqrt{n^2 + 1}} \right].
\label{eq:nonosc}
\end{equation}
According to Eqs. (\ref{eq:resdensity}, \ref{eq:deltaro}, 
\ref{eq:nscale})
\begin{equation}
\frac{n}{\sqrt{n^2 + 1}} = - \cos 2\theta_m.
\label{eq:nrelat}
\end{equation}
The maximal possible  $|n|$ corresponds to the minimal value of angle, that is 
$\theta_m =\theta$, 
which gives $P =  \sin^2 \theta$.
Since 
\begin{equation}
\frac{n_0}{\sqrt{n_0^2 + 1}} = - \cos 2 \theta_0, ~~~~
\frac{n}{\sqrt{n^2 + 1}} = - \cos 2 \theta,  
\label{eq:cosrelation}
\end{equation}
Eq. (\ref{eq:adsolution}) can be rewritten as  
\begin{equation}
P = \frac{1}{2}\left[1 + 
\cos 2\theta_0 \cos 2\theta \right],  
\label{eq:usualform}
\end{equation}
which coincides with the well know now expression. 

To avoid problems with publications, we tried to hide the term resonance, 
 did not discuss solar neutrinos and even 
did not include the reference to our paper on the resonance 
enhancement. This did not help. The paper submitted 
in the fall of 1985 to JETP Letters, 
was rejected with motivation: no reason for quick 
publication. It was resubmitted to JETP in December 1985. 
The results  were reported  at the 6th Moriond workshop  (January 1986). 
The paper was reprinted in  the ``Solar neutrinos: 
the first Thirty Years" \cite{30year}.
Latter I reproduced the English translation of 
the paper and posted it on the arXiv  \cite{arx}. 

In the ``Perestroyka'' time  the rule was introduced that a 
paper can be submitted to 
a journal abroad only after results have alredy been  published in Soviet journal. 
That was enormously long procedure, so we decided to present our results at conferences  
and then put all the material in reviews \cite{uspechi}.

\subsection{Wolfenstein's unknown paper}
%%%%%%%%%%%%%%%%%%%%%%%%%%%%%%%%%%%%%%%%%%%%%%%%%%%%

In the same 1978 Wolfenstein wrote the paper 
``Effect of matter on Neutrino oscillations'' (the same title  as for \cite{w79a})
published as contributed paper in the proceedings 
of  ``Neutrino-78'' \cite{nu78}. 
Wolfenstein mentioned this paper in his letter to us. 
We thought that this is just a conference version 
of what had already been in the published paper \cite{w78a}.   
The contribution~\cite{nu78} (not even a talk) 
has practically no citations and no impact. 
I read the paper for the first time in 2003, 
after  E. Lisi asked me to send a copy of the paper   
which he saw in the ICTP library. 
%%check if proceedings 
%%of the Purdue conference are available in the ICTP library.  
%%They were! 
The content of the paper is amazing 
and it is not clear why Wolfenstein did not publish it in any journal.

Wolfenstein still considered the case of massless neutrinos. 
He noticed that in the Sun the mixing in matter varies due 
to change of the chemical composition (if the couplings $g_n$ and $g_p$ 
in (\ref{eq:currentscat}) are different).  
The ratio $y \equiv$ neutron/proton decreases from 0.41 
in the center to 0.13 at the surface.    
For constant $y$ the mixing would be constant for massless 
neutrinos in spite of strong total density change. 
In the original paper \cite{w78a} he neglected this change 
and considered constant averaged density.  
Wolfenstein writes in~\cite{nu78}:
``The percentage change in $\theta_m$ per oscillation 
is small (since there are  1000 oscillations 
on the way out the sun) [A.S.: this is the adiabatic condition],  
so that we can apply the adiabatic approximation''.  
And then he gives the formula without any derivations, 
explanations or comments:
\begin{equation}
|\langle \nu_e | \nu_e (x)\rangle |^2 =
\cos^2\theta_0 \cos^2\theta_m(x) + 
\sin^2\theta_0 \sin^2\theta_m(x)  
+ 0.5 \sin 2\theta_0 \sin 2\theta_m(x) \cos \Phi(x),    
\label{eq:adformw}
\end{equation}
where $\theta_0$ and $\theta_m(x)$ are the mixing angles in matter 
in the production and in a given point $x$.  
He concludes: ``In this case [AS: varying effective density] neutrinos are transformed 
not only by virtue of the oscillating phase 
but also by adiabatic change in propagating eigenvectors.'' 
[A bit obscure but now we understand the meaning.] 
``For example, if $\theta_0 = 0$, the oscillating term 
vanishes but there is transformation of $\nu_e$ 
into $\nu_\mu$ since neutrino is propagating in eigenstate 
which originally $\nu_e$ but adiabatically transforms 
into a mixture of $\nu_e$  and $\nu_\mu$''.  
This is nothing but description of the adiabatic conversion! 
%which can be translated in our present language in the following way. 

%%\subsection{Publications}
%%%%%%%%%%%%%%%%%%%%%%%%%%%%%%%%%%%%%%%%%%%%%%%%%%%%%%

%%%%%%%%%%%%%%%%%%%%%%%%%%%%%%%%%%%%%%%%%%%%%%%%%%%%%%%%%%
\section{Further developments 1985 - 1986}
%%%%%%%%%%%%%%%%%%%%%%%%%%%%%%%%%%%%%%%%%%%%%%%%%%%%%%%%%%%%

\subsection{MSW as the level crossing phenomenon}
%%%%%%%%%%%%%%%%%%%%%%%%%%%%%%%%%%%%%%%%%%%%%%%%%

\begin{figure}
\begin{center}
\begin{minipage}{0.33\linewidth}
\centerline{\includegraphics[width=1.0\linewidth]{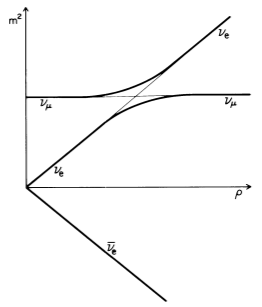}}
\end{minipage}
\end{center}
\caption[]{The level crossing scheme for $2\nu$ case. 
Dependence of the eigenvalues of the Hamiltonian in matter 
(effective masses) on density. From  \cite{bethe}.}
\label{fig:levelcr}
\end{figure}

As far as I know, N. Cabibbo did not publish his WIN-85 summary talk with  
description of the MSW effect in terms of the level crossing, 
Fig.~\ref{fig:levelcr}.   
We did not proceed in this direction either. 
Description in terms of the {\it eigenvalues}  
of the system (levels) is complementary to ours
and  I was happy with the description in terms of the {\it eigenstates}. 
 
Half a year after WIN-85, apparently not knowing about 
Cabibbo talk, H. A. Bethe considered the dependence 
of the eigenvalues of the Hamiltonian in matter (effective masses) 
on density \cite{bethe}.  He noticed that minimal splitting 
is in  resonance. The adiabatic evolution appears as motion 
of the system along a given fixed 
level without jump to another level (Fig.~\ref{fig:levelcr}). Thus, $\nu_e$ produced 
at high density follows the upper curve, which is equivalent 
to the absence of transition between the eigenstates. 
This presentation was extremely important for further developments. 

%%%%%%%%%%%ffff9%%%%%%%%%%%%%%%%%%%%%%%%%%%%%%%%%%%%%%%%%%%%%%%%%%%%%%%%%%%%
%%%%%%%%%%%%%%%%%%%%%%%%%%%%%%%%%%%%%%%%%%%%%%%%%%%%%%%%%%%%%%%%%%%%%%%%%%

By the way, one can find almost complete level crossing scheme in \cite{barger80}, 
where the dependence of the eigenvalues 
on energy (not density) was presented for constant density.  

\subsection{1986 Moriond workshop}
%%%%%%%%%%%%%%%%%%%%%%%%%%%%%%%%%%%%%%%%%%%%%%%%%%%%%%%%%%%%

I was  invited to the 6th Moriond workshop in Tignes (January 25 - February 1, 1986)
again by recommendation of A. Pomansky. I had suggested a talk 
``Neutrino oscillations in matter with varying density''  
and the organizers gave me 15 min in one of the evening sessions 
on the second or third day.  I was very much surprised when 
in the bus from Bourg-Saint-Maurice to Tignes 
somebody told me that Peter Rosen (who gave the introductory talk) wants to talk to me, 
and the talk will be, to a large extent,  on the ``Mikheyev-Smirnov" mechanism.  
Things changed quickly after arrival:
Jean Tran Thanh Van told me that the organizers arranged my 30 - 40 min talk  
and,  if needed, will find time for additional presentations. Eventually, I gave  
two talks which  
%% (I had no time to prepare more transparencies). 
included the theory of resonance oscillations and conversion 
in varying density medium, graphical representation, 
applications to supernova neutrinos 
(covering the material of \cite{msjetp}), effects 
in the Earth for atmospheric, supernova and solar neutrinos, 
transformations in the Early Universe.  
%%I did not spoken  on conversion of solar neutrinos in the Sun. 

%%I also gave a talk (by request from collaboration) 
%%on Baksan underground data on Cygnus X-3. 

\subsection{Completing theory of adiabatic conversion}
%%%%%%%%%%%%%%%%%%%%%%%%%%%%%%%%%%%%%%%%%%%%%%%%%%%%%%%%%%%

At the same Moriond  workshop A. Messiah gave  a talk ``Treatment 
of $\nu_{sun}$ -oscillations in solar matter. The MSW effect'' 
\cite{messiah}. In the proceedings he writes 
``MS call it the resonant amplification effect - 
a somewhat misleading denomination".  
He did not like and did not use the term ``resonance'', 
claiming that the effect can be readily deduced 
from the adiabatic solution of the equation of 
flavor evolution.    
During the  workshop Messiah told me: ``Why do you call your effect 
resonance oscillations? This is confusing, I will call it the MSW effect". 
I agreed (see Fig. \ref{fig:adconversion},  and whole the story in ``Solar neutrinos: 
Oscillations or no-oscillations" \cite{Smirnov:2016xzf}). 
That was the first time I heard a term ``the MSW effect".

Messiah  used complicated notations, operator forms, {\it etc.}
He derived the equation for the evolution matrix 
$U_H (x, x')$ of the eigenstates in matter:  
\begin{equation}
\frac{dU_H(x, x')}{dx}  = 
\left[ W(x) {\bf \sigma} {\bf w}(x')
+ \frac{d \theta_m}{dx} {\bf \sigma} {\bf k}
\right] U_H(x, x'),    
\label{eq:adiabmess}
\end{equation}
where $W(x)$ is the level spacing. 
Translating the content of this equation to the language/notations 
we use now,  one finds that  ${\bf \sigma} {\bf w}(x')$
is essentially $\sigma_3$,  ${\bf \sigma} {\bf k}$ is 
$\sigma_2$, {\it etc.} If one also  uses $\nu_m(x)$ instead of $U_H(x, x')$,  
Eq. (\ref{eq:adiabmess}) is reduced to the well known now 
equation for the eigenstates $\nu_{im}$. 
The derivative $d \theta_m/dx$ appears in the evolution equation for the first time. The  
adiabatic solution is obtained when the second term on the RHS 
can be neglected. This gives the adiabatic condition 
\begin{equation}
\omega \equiv  \frac{d \theta_m/dx }{2W(x)} \ll 1. 
\label{eq:addcond}
\end{equation}
The essence of the adiabatic parameter $\omega$ is 
$$
\omega = \frac{\rm Rotation~ velocity~ of~ 
eigenvector}{\rm level~ spacing}.  
$$
``If the eigenvectors of the Hamiltonian in matter rotate slowly,  
the components (projections) of the vector of neutrino state 
along the rotating eigenvectors stay constant".

Messiah discussed the adiabaticity violation.  
He introduced the amplitude of transition between the eigenstates 
$\beta \equiv A(\nu_{2m} \rightarrow \nu_{1m})$ ($\nu_{1m} = \nu_1$ for final state 
in vacuum), which quantifies adiabaticity violation. He obtained the formula   
\begin{equation} 
P_{ee} = \frac{1}{2} [1 + (1 - 2|\beta|^2) 
\cos2\theta  \cos2 \theta_m(x')] 
\label{eq:parkef}
\end{equation}
called later the Parke's formula. Here $|\beta|^2 = P_c$ is 
the jump or flop probability. 
The adiabatic solution corresponds to $\beta = 0$.
Massiah considered explicitly weak violation of the adiabaticity and found that   
corrections to the adiabatic solution are proportional to $\omega^2$.

The result of adiabatic evolution from high densities to vacuum 
can be  written in terms of the neutrino states~\cite{messiah}:
\begin{equation}
|\nu_{final} \rangle = \cos \theta_m^0 e^{i\phi_1} |\nu_1 \rangle +
\sin \theta_m^0 e^{i\phi_2} |\nu_2 \rangle ,   
\label{eq:adsolution1}
\end{equation}
where $\theta_m^0$ is the mixing angle in matter in the production point.
%%The suppression pit for solar neutrinos has been computed. 

%%%%%%%%%%%%%%%%%%%%%%%%%%%%%%%%%%%%%%%%%%%%%%%%%%%%%%%%
\subsection{Adiabaticity violation - non-perturbative result}
%%%%%%%%%%%%%%%%%%%%%%%%%%%%%%%%%%%%%%%%%%%%%%%%

W. Haxton \cite{haxton} and S. Parke \cite{parke} considered the  
adiabaticity violation using the 
level crossing picture in analogy to the level crossing 
problem in atomic physics. They showed that transitions between the levels 
can be  described by the Landau - Zenner probability valid for 
linear dependence of density on distance. 
In this case the flop probability equals  
\begin{equation}
P_c = |\beta|^2 = P_{LZ} = e^{- \pi \gamma/2},                 
\label{eq:lzprob}
\end{equation}
where $\gamma$ is the adiabaticity parameter:    
$\gamma = 2\pi \kappa_R$,  and $\kappa_R$ was defined in 
(\ref{eq:adiabpar}).

W. Haxton \cite{haxton} focussed on the ``non-adiabatic 
solution" for solar neutrinos. Along the diagonal side  of the MSW triangle 
in the ($\Delta m^2 - \sin^2 2\theta$) plane   
with  initial $\theta_m \approx \pi/2$  and final $\theta \approx 0$ 
the survival probability equals  
$$
P_{ee} \approx P_c.                    
$$

S. Parke~\cite{parke} derived general expression for the survival probability 
for $\theta_m \neq \pi/2$ and non-zero $\theta$, which is  
the same expression as in the Messiah's paper (\ref{eq:parkef}).     
%%I should say that I learned these results  from Parke  
%%and not Messiah paper which was difficult to understand. 

\subsection{Graphic approach}
%%%%%%%%%%%%%%%%%%%%%%%%%%%%%%%%%%%%%%

According to S. Petcov, in his talk in Savonlinna N. Cabibbo also used graphic representation which,  
it seems,  differs from the ours. 
%%Cabibbo did not referred to my talk. 
In his picture  flavor axis was $\nu_e$ in the up direction and $\nu_\mu$ is in the down direction.  
That should be the representation for probabilities, {\it i.e.} for  the neutrino polarization vector.  
%%Oscillations are  analogous  to the electron spin-precession in the magnetic field. 

The graphic representation of conversion which uses the analogy with precession 
of the electron spin in the magnetic field appeared in the paper~\cite{bouchez}. 
The neutrino polarization vector has components 
$(E, Y, X)$  which coincide with our  $(P, R, I)$ (\ref{eq:pridef}). 
The vector precesses around the Hamiltonian (magnetic field) axis
$\vec{B} = 2\pi/l_m (\sin 2\theta_m,~ 0,~ \cos 2\theta_m )$. 
%%Evolution of neutrinos is equivalent to  
%%precession of vector on the surface of the cone. 
Adiabatic conversion is driven by the rotation of the cone axis (Hamiltonian). 
Specific example was considered with 
$(d \theta_m/dx)/2W = const$ which is analytically solvable. 
%%Figure
 
%%%%%%%%%%%ffff10%%%%%%%%%%%%%%%%%%%%%%%%%%%%%%%%%%%%%%%%%%%%%%%%%%%%%%%%%%%%
%%\begin{figure}
%%\begin{center}
%%\begin{minipage}{0.33\linewidth}
%%\centerline{\includegraphics[width=0.7\linewidth,draft=true]{bethe.png}}
%%\end{minipage}
%%\end{center}
%%\caption[]{Graphic representation in terms of probabilities. From~\cite{bouchez}}
%%\label{fig:grap-pro}
%%\end{figure}
%%%%%%%%%%%%%%%%%%%%%%%%%%%%%%%%%%%%%%%%%%%%%%%%%%%%%%%%%%%%%%%%%%%%%%%%%%

Among other things the authors of~\cite{bouchez}  write that ``As we shall see this 
(resonance oscillations) is not exactly 
what happens in the Sun (varying density)''.  
See a discussion of this point in \cite{Smirnov:2016xzf}.

%%\subsection{Further developments}
%%%%%%%%%%%%%%%%%%%%%%%%%%%%%%%%%%%%%%%%%%

\section{Summary and Epilogue}
%%%%%%%%%%%%%%%%%%%%%%%%%%%%%%%%%%%%%%%%%%%%%%%

I would summarize the $W$ and $MS$ contributions to the MSW effect in the following way.

{\it Wolfenstein:}  Coherent forward scattering should be taken 
into account. It induces oscillations of massless neutrinos  and modifies    
 oscillations of massive neutrinos.  Strong  modification of oscillations appears at  
$l_\nu  \sim   l_0$;  transition probability can be enhanced. 
The evolution equation is derived.   
In a largely unknown paper, the adiabaticity condition for massless neutrinos is formulated qualitatively   
and adiabatic formula for probability is presented.  

Barger, Whisnant, Pakvasa and Phillips marked the condition of maximal mixing in matter and 
explored enhancement (as well as suppression) of oscillations. 

{\it Mikheyev-Smirnov:} Resonance phenomenon was uncovered, 
the properties of the resonance and resonance enhancement of oscillations were  
studied. The adiabatic condition was formulated and quantified and adiabatic 
transitions for massive neutrinos  described. Equation for components of the neutrino 
flavor polarization vector 
was derived. Graphic representation was elaborated.

{\it Further developments}  made by  N. Cabibbo, H. Bethe, A. Messiah, W. Haxton, S. Parke. 
 MSW was interpreted as the level crossing phenomenon. Adiabaticity violation formalism was developed, 
flop or jump probability -- computed. \\

After the initial period, the   studies  proceeded in various directions which include 
(i) dynamics of the conversion and  theory of adiabaticity violation; 
(ii) conversion in media with different properties: 
thermal, polarized,  magnetized, moving, periodic, fluctuating, stochastic;  
%%Mixing induced by matter (due to NSI) has been further studied, 
(iii) spin-flavor conversion in magnetic fields; (iv) effects in different neutrino (antineutrino) channels; 
(v) numerous applications. \\

%%\section{Epilog}
%%%%%%%%%%%%%%%%%%%%%%%%%%%%%%%%%%%%%%%%%%%%%%%%%%%

1998  In final Homestake  publication  
there is no even reference to the MSW solutions. Neutrino spin-flip in magnetic 
field was  considered as the main explanation.

2002 -  2004 LMA MSW was established by SNO and KamLAND  
as the solution of the solar neutrino problem.  

2008 N. Cabibbo: data confirmed the original Pontecorvo 
proposal for the solution of  the solar neutrino problem and rejected  
the ``spurious  MSW  solution". 

2015 In the scientific background description the 
Nobel prize committee presented formula for oscillations 
in medium with constant density in connection to the solar neutrinos. (See~\cite{Smirnov:2016xzf}.) 

2017 BOREXINO further confirmed the LMA MSW solution. 

2018 K. Lande: Homestake did not observed time variations of signal.

%%There are almost unknown papers in which some results have been obtained for the first time 
%%there are later paper with enormous impact in which similar results have been obtained.   

\section*{Acknowledgments}

I am grateful to S. T. Petcov and D. Wyler for providing me valuable historical information  
and to E. K. Akhmedov for discussions.  

%%\section*{Appendix}

%% We can insert an appendix here and place equations so that they are

%%%%%%%%%%%%%%%%%%%%%%%%%%%%%%%%%%%%%%%%%%%%%%%%%%%%%%%%%%%%%%%%%%%%%%%%%%%%%%%%%%%%%

\end{document}